\documentclass[sigplan,screen]{acmart}

\renewcommand\footnotetextcopyrightpermission[1]{} 
\AtBeginDocument{%
	\providecommand\BibTeX{{%
			\normalfont B\kern-0.5em{\scshape i\kern-0.25em b}\kern-0.8em\TeX}}}

\usepackage{multirow}
\usepackage{makecell}

\begin{document}
	
	\title{An Empirical Study on Fertility Proposals Using Multi-Grained Topic Analysis Methods}

	\author{Yuilin Zhou}
	\email{bluedyson@stu.haut.edu.cn}
	\affiliation{%
		\institution{Henan university of technology}
		\city{ZhengZhou}
		\state{Henan}
		\country{China}
	}

	\begin{abstract}
		Fertility issues are closely related to population security, in 60 years China's population for the first time in a negative growth trend, the change of fertility policy is of great concern to the community. 2023 ``two sessions" proposal ``suggests that the country in the form of legislation, the birth of the registration of the cancellation of the marriage restriction" This topic was once a hot topic on the Internet, and ``unbundling" the relationship between birth registration and marriage has become the focus of social debate. In this paper, we adopt co-occurrence semantic analysis, topic analysis and sentiment analysis to conduct multi-granularity semantic analysis of microblog comments. It is found that the discussion on the proposal of ``removing marriage restrictions from birth registration" involves the individual, society and the state at three dimensions, and is detailed into social issues such as personal behaviour, social ethics and law, and national policy, with people's sentiment inclined to be negative in most of the topics. Based on this, eight proposals were made to provide a reference for governmental decision making and to form a reference method for researching public opinion on political issues.
	\end{abstract}

	
	\begin{teaserfigure}
		\includegraphics[width=\textwidth]{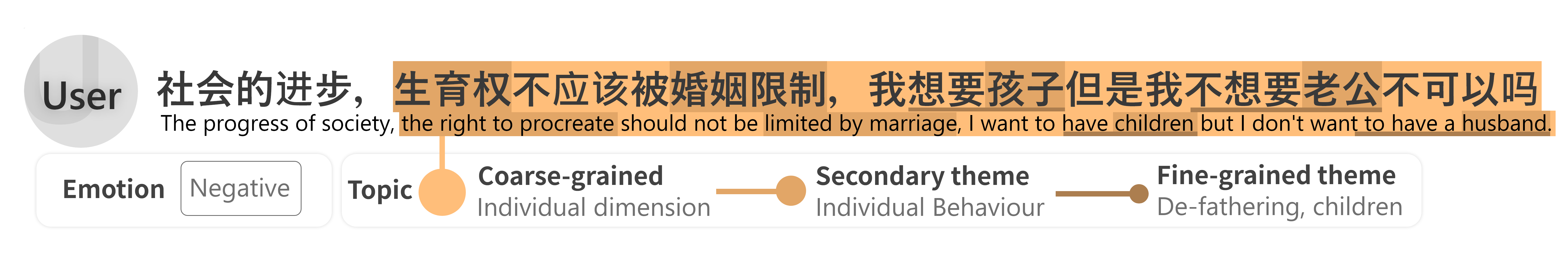}
		\caption{A result from multi-granularity semantic analysis framework}
		\label{fig:teaser}
	\end{teaserfigure}

	\maketitle
	
	\section{Introduction}
	General Secretary Xi Jinping emphasised that the population issue has always been a global, long-term and strategic issue for China. The National Bureau of Statistics released data showing that the national population at the end of 2022 was 141.175 million, a decrease of 850,000 from the end of the previous year. Annual births of 9.56 million people, the birth rate of 6.77‰; deaths of 10.41 million people, the population mortality rate of 7.37‰; natural population growth rate of -0.60‰. Fertility issues are closely related to population security, and with China's population in negative growth for the first time in 60 years, changes in fertility policy are of great concern to society. Shifting the focus of birth registration to the level of fertility intentions and birth outcomes is an inherent requirement for promoting the implementation of family planning policies.
	
	During the 2023 ``Two Sessions", Xie Wenmin, a member of the National Committee of the Chinese People's Political Consultative Conference (CPPCC), proposed that marriage restrictions should be removed from the birth registry in the form of legislation. The topic hit the Sina microblogging site's Hot Search on 28 February. The ``unbundling" of birth registration and marriage became the focus of heated social debate.
	
	In the light of the discussion triggered by the proposal of the ``Two Sessions", this study focuses on the social issues involved in the discussion of the ``unbundling" of birth registration and marriage, as well as the attitudes and positions of the public on this proposal.
	
	\section{Related work}
	The rapid development of social network platforms represented by Weibo provides channels for people to express their opinions and participate in politics. \cite{001} Public opinion, as an object with close connection with governmental decision-making, Li Ting and other scholars believe that on the issue of fertility, the formulation of effective policies around the specific reasons shown by public opinion can only promote the rebound of the fertility rate. And by studying the public opinion related to the opening up of the two-child policy on the Weibo, it is found that the public will take the initiative to state the objective limitations faced by childbirth, and the attitude of the public tends to be negative, and reflects that the inhibiting effect of the real conditions is the biggest reason for hindering the rebound of the fertility rate. \cite{002} In further studies by other scholars, it is found that the public's reasons are becoming more diversified and detailed, ranging from the issue of women's individual rights and interests to the property policy \cite{003}, pension security and other ever-detailed areas\cite{004}\cite{005} . Therefore, in the face of diversified and detailed opinions, a refined big data analysis can completely understand public opinion and provide public support for further improving maternity protection. Meanwhile, most of the studies in this area focus on public opinion after the implementation of the policy, and few of them are concerned with public opinion during the consultation period, so this study mainly analyses the latter.
	
	Constructing semantic co-occurrence network is an effective way to reflect the deep structural relationship of text content and clarify the relationship between the topics of text content. \cite{006} Li Lei and other scholars introduced the association clustering algorithm into the topic identification of online public opinion information in the co-occurrence network, and achieved better results. \cite{007} Li Gang and other scholars further introduced Louvain community discovery algorithm to further optimize the overall topic recognition of network public opinion. \cite{008} In summary, doing community discovery by constructing semantic co-occurrence networks for online comment texts can understand discussion topics at a coarser granularity, while finer-grained topics need to be further mined using topic models.
	
	Topic models have the ability to mine the mainstream opinions in public opinion \cite{009}, but the feature words generated by traditional LDA (latent dirichlet allocation) models lack semantic interpretability \cite{010}, rely on manually setting hyperparameters such as the number of topics, which makes the algorithms less robust, and are not effective in short texts such as microblogs \cite{011}. Using the Bertopic framework composed based on pre-trained models such as Bert, it is better than LDA in terms of language comprehension and generation ability \cite{012} and algorithm robustness \cite{013}, and it can more accurately mine the topic classification in the face of short texts \cite{014}. Meanwhile, in the experiments of CA-LDA model proposed by Cai and Changqing, it is found that embedding semantic co-occurrence network to do community discovery results as a priori probability can effectively improve the LDA topic classification results under short text. \cite{015} In this study, we will use the semantic co-occurrence network results as the a priori probability clustering of the Bertopic model to obtain the topic classification results, and further use hierarchical clustering to obtain multi-granularity results, to clarify the direction of the majority of the audience's concern, the content of some of them, and the details of the specific audience's concern in the different audience ranges. And then provide the government with the basis of public opinion, and to generalise the formation of the reference of the research of the public opinion of the political issues. Lastly, it suggests a way of researching political issues.
	
	\section{Data collection and analysis}
	In this study, Sina Weibo was selected as the data collection platform, and the proposal to abolish marriage restrictions on birth registration was used as a representative policy issue for quantitative research. We captured the comments under the topic of ``Suggesting the cancellation of marriage restrictions on birth registration", and obtained 12,509 comments after de-weighting and cleaning the data. Semantic co-occurrence network analysis is used to explore the domain of comment discussion, and the keyword results are fed into the Bertopic model as a priori experience. The resulting fine-grained topics are added to the sentiment analysis results and then hierarchically clustered to form the multi-granularity results.
	
	\begin{figure}[h]
		\centering
		\includegraphics[width=\linewidth]{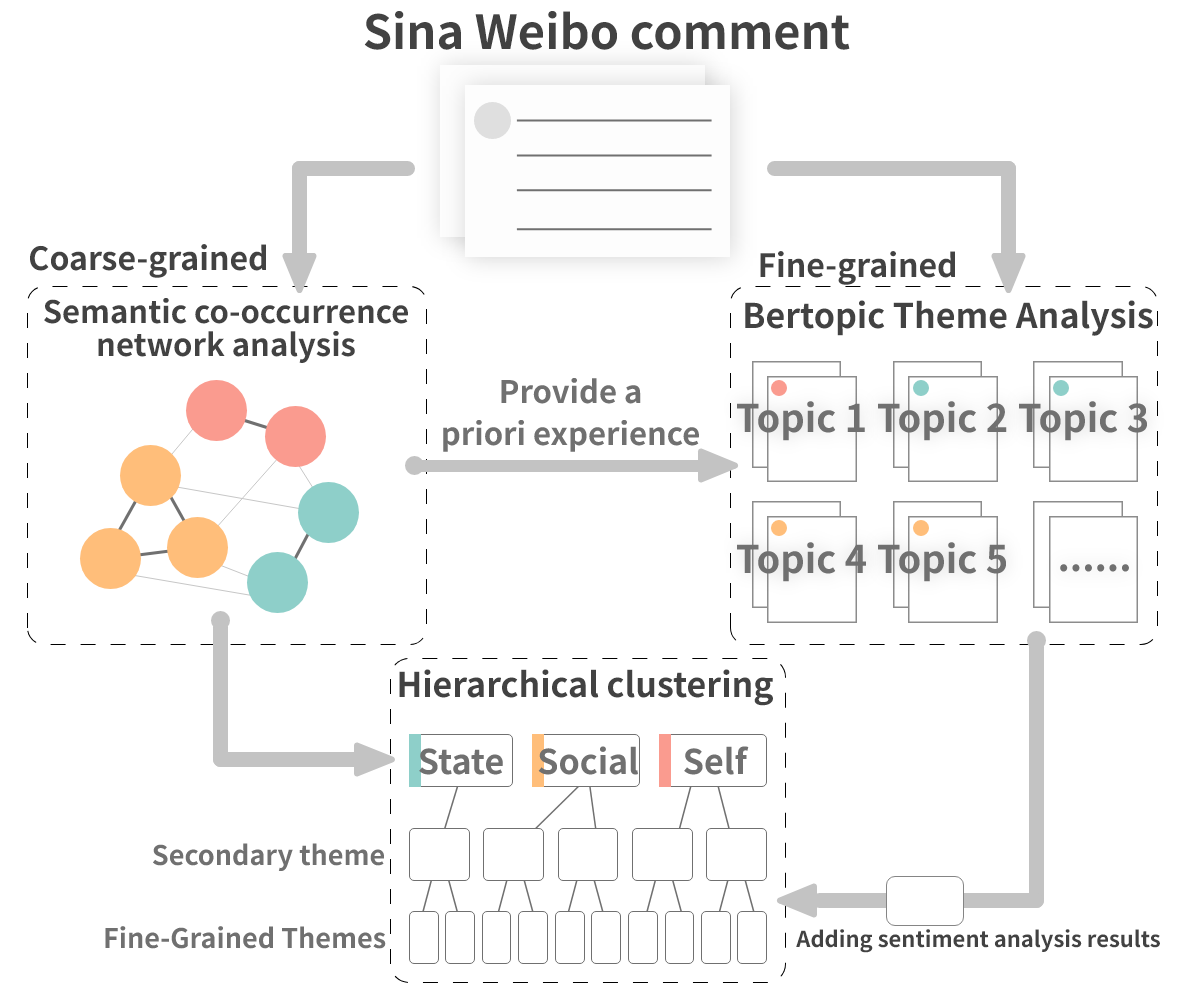}
		\caption{Multi-granularity semantic analysis framework}
	\end{figure}
	
	\subsection{Discussion area}
	
	\begin{figure}[h]
		\centering
		\includegraphics[width=\linewidth]{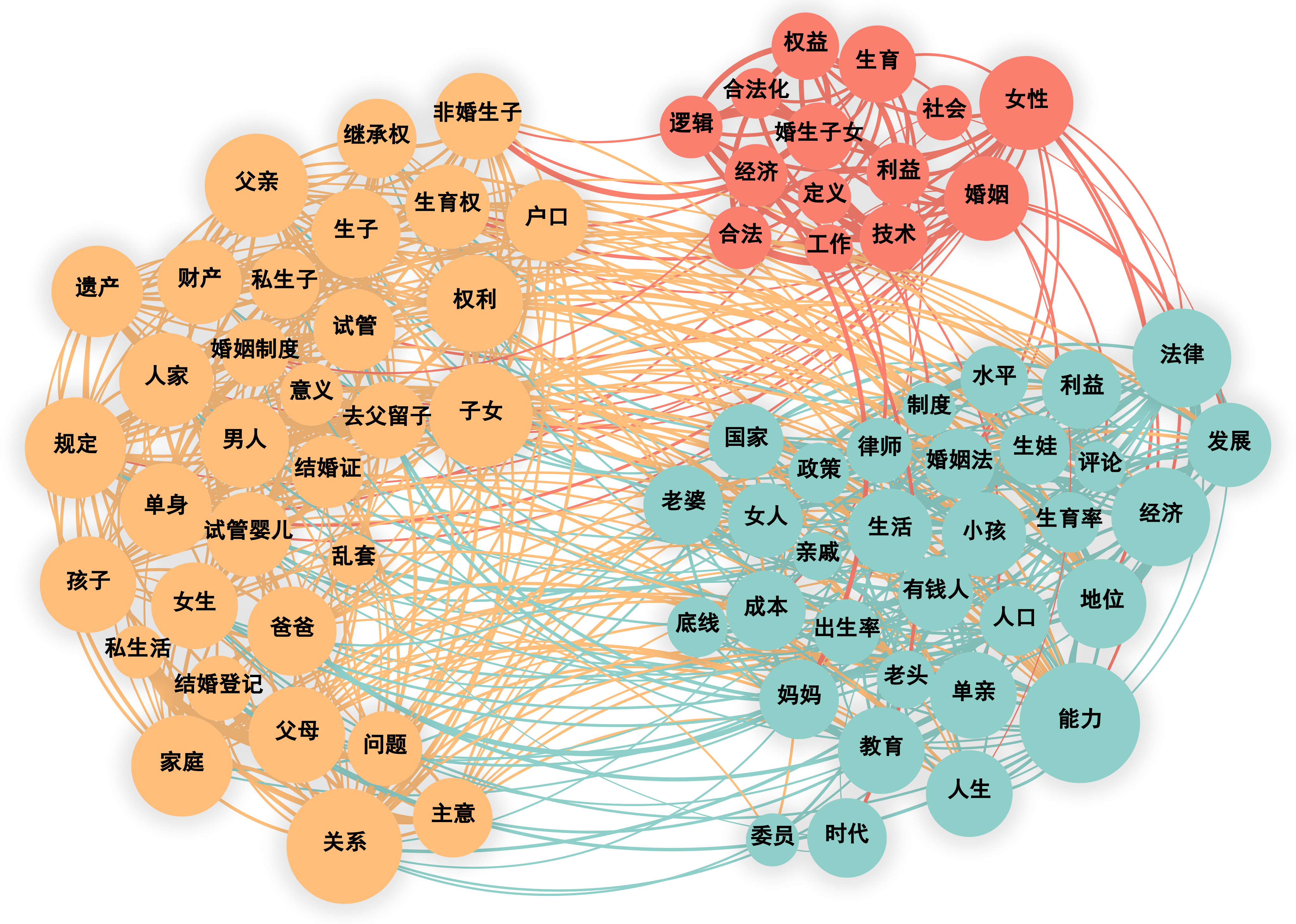}
		\caption{Semantic co-occurrence network for microblog comments}
	\end{figure}
	
	In this study, the discussion domains of popular discussions are explored by finding discussion communities through semantic co-occurrence networks. Defining the unit of analysis as a single microblog comment, the probability of two words appearing in the same comment is the numerator, and the product of the probability of two words appearing in the comment separately is the denominator. The ratio is the correlation between the two words in the comment, greater than 1 when the correlation between the words is high, with the significance of the analysis.
	
	The constructed semantic co-occurrence network is put into Gephi, and the Louvain community discovery algorithm is used to search for user discussion domains, and three community classifications are obtained using a resolution of 1.2, and it has a modularity degree of 0.347, which can be regarded as a high degree of independence between communities and a reasonable division. By observing the vocabulary distribution of the co-occurrence network, it can be found that the public is discussing three levels: national, social, and personal.
	
	As shown in figure 3, the public can be found discussing the three communities of the State, society and the individual, accounting for 43.24\%, 41.89\% and 14.86\% respectively. \textbf{The first community} is at the individual level, focusing on the issues of individuals and the families in which they live, with \textit{``children, fathers, women, and men"} being the identity of individuals in society, \textit{``rights, reproductive rights, and inheritance rights"} relating to the rights and interests of individuals, and \textit{``family, inheritance, property, parents, and children"} stressing the importance of the family, property, parents, and children. \textit{``Family, inheritance, property, parents, children"} emphasise inter-individual issues within the family sphere. \textbf{The second community} is at the national level, focusing on laws and policies, and the legal and policy issues of marriage are mentioned, with emphasis on issues such as ``childbearing, education", etc. The third community is at the social level, emphasising women's rights and interests. \textbf{The third community} is at the social level, emphasising the social realities faced by women, such as \textit{``work, childbirth, marriage and rights and interests"}. The three communities are interrelated, with strong commonality in discourse understanding and semantic expression. Among them, the first and second communities are more connected, indicating that the institutions and laws related to childbirth and marriage are inseparable from personal development, and are the topics that netizens focus on. As an important social subject, the issue of women's rights and interests at the social level has become an important influence on netizens' attitudes towards the proposal. In addition, ethical and moral issues run through the three communities, with the individual level manifesting itself in the family ethics on paternity; the social level manifesting itself in the social and ethical dimensions of technical reproduction and matrimonial regimes and the moral dimension of the protection of women's rights and interests; and the national level manifesting itself in the extension of the individual and social dimensions, which involves the corresponding legal and institutional issues.
	
	\subsection{Fine-grained thematic analysis}
	For sentiment analysis, this study uses three categories of positive, negative, and no sentiment for classification, and the model selects ``RoBERTa-wwm-ext, Chinese" \cite{016} from Xunfei Laboratory of Harbin Institute of Technology, and is trained using template cueing learning on the artificially reclassified SMP2020-EWECT microblog sentiment classification data set and trained using template cue learning. The accuracy rate is 91\% on the test set after multiple rounds of training, which is at a high level and the model has high confidence. The following sentiment analysis results in this study are all derived from the above model.
	
	For topic analysis, this study adopts the Bertopic framework for a multi-granularity topic clustering algorithm with guidelines. Three community keywords distinguished by co-occurring networks as a priori information are embedded as guidelines in the comments matching similar topics to obtain more interpretable clustering results. The same RoBERTa model as above is used as a text feature extractor to extract features from the comments with mixed a priori information, which is further processed using dimensionality reduction and clustering algorithms to obtain the clustering results, and based on the results, hierarchical clustering is used to summarise the results.

	\begin{table*}[]
		\footnotesize
		\caption{Multi-granularity topic keyword results}
		\begin{tabular}{llll}
			\hline
			\begin{tabular}[c]{@{}l@{}}co-occurrence \\ network analysis\end{tabular} & Secondary topic & Subject Keywords                                                                                                                                   & Fine-grained subject keywords                                                                                                                                                                                                                                                         \\ \hline
			Individual dimension                                                      & 1 - Behaviour   & \begin{tabular}[c]{@{}l@{}}Children, women, men, single, \\ out of wedlock,marriage licence,\\ life\end{tabular}                                   & \begin{tabular}[c]{@{}l@{}}Fertility, de-fathering, childbearing, singleness, \\ will, children, population, eugenics\\    \\ Chaos, human nature, cursing, human traffickers, \\ trafficking in women, rape, human trafficking, \\ monsters\end{tabular}                             \\ \cline{4-4} 
			& 5-Ethics        & \begin{tabular}[c]{@{}l@{}}Illegitimate children, children born out of wedlock, \\ inheritance, property, children, rights, education\end{tabular} & \begin{tabular}[c]{@{}l@{}}children, entitlement, parenting, inheritance, \\ costs, issues, mother's name, legal\\    \\ Bottom line, challenge, shackles, single, top, \\ entitlement, extramarital affairs, vested interests\end{tabular}                                           \\ \cline{4-4} 
			Social dimension                                                          & 2-Ethics        & \begin{tabular}[c]{@{}l@{}}Logic, Meaning, Bottom Line, Habits,\\ Superiority, Private Life, Restraints\end{tabular}                               & \begin{tabular}[c]{@{}l@{}}Marriage, marriage registration, tradition, \\ extramarital affairs, finding a partner, couples, \\ moral hazard, life\\    \\ Childbirth, women, illegitimate children,\\ children out of wedlock, children, society, \\ bottom line, issues\end{tabular} \\ \cline{4-4} 
			& 3 - Laws        & \begin{tabular}[c]{@{}l@{}}Legalisation, definition, law, etiquette,\\ hooliganism, legitimate interests, approach\end{tabular}                    & \begin{tabular}[c]{@{}l@{}}Fertility, IVF, Single Parents, Women, \\ Problems, Advice, Education, Costs\\    \\ Marriage Law, Policy, Marriage Licence, Inheritance,\\ Marriage Registration, Property, Inheritance, Legal\end{tabular}                                               \\ \cline{4-4} 
			National dimension                                                        & 4-State         & \begin{tabular}[c]{@{}l@{}}Recommendations, population, country,\\ marriage regime, fertility, dividend, birth rate\end{tabular}                   & \begin{tabular}[c]{@{}l@{}}illegitimate child, inheritance, illegitimate child,\\ marriage, half-brother, paternity test, policy, \\ household registration\\    \\ Experts, commissioners, quality, fear of parenthood,\\ culture, bride price, buying a house, times\end{tabular}   \\ \hline
		\end{tabular}
	\end{table*}
	
	In the discussion of policy topics, the secondary topic is the main discussion direction of the topic, while the discussion content belongs to the topic is the further refinement of the direction and specific content to form the above Table 1. Among them, the discussion at the individual level mainly revolves around the Topic1-Personal Behaviour and Topic5-Family Ethics, the Topic1 discussion mainly revolves around the public order, morality and ethics of the individual, and the Topic5 discussion revolves around aspects such as the foetus's right to health and civil rights; at the social level, the public mainly revolves around the aspects of Topic2-Moral Ethics and Topic3-Laws and Regulations. Discussions on socially fair ethical norms of marital relations are concentrated in Topic2, while discussions on the reasonableness of the proposal itself, its relevance to existing relevant laws, and its relationship with existing highly controversial laws and regulations are concentrated in Topic3; public discussions on the national level are mainly centred on Topic4-policy itself, and the public is more concerned about the impact of policy changes on their own lives under the pressure of practical factors and other factors. Under the pressure of practical factors and other aspects, the public is more concerned about the impact of policy changes on their own lives.
	
	Cross-analysing sentiment and topics (Figure 4) reveals a high percentage of negative sentiment under all topics. Under Topic1-Personal Behaviour and Topic4-National Policies, the negative sentiment is more than 55\%, and the public's negative emotional attitude is significant. Among the other three topics, only Topic2-Moral and Ethical has a negative affective attitude of less than 45\%. Among them, only under Topic 4-National Policy, the proportion of positive emotions is lower than 10\%, while the remaining four topics are all higher than 10\%, and the proportion of positive emotions in Topic 5-Family Ethics reaches 16.83\%, which is much higher than that of the other topics.
	
	The sources of negative public sentiment were analysed by combining the above thematic analysis keywords, the results of the sentiment analysis and some of the blog posts, which in turn led to the conclusions below.
	
	\begin{figure}[h]
		\centering
		\includegraphics[width=\linewidth]{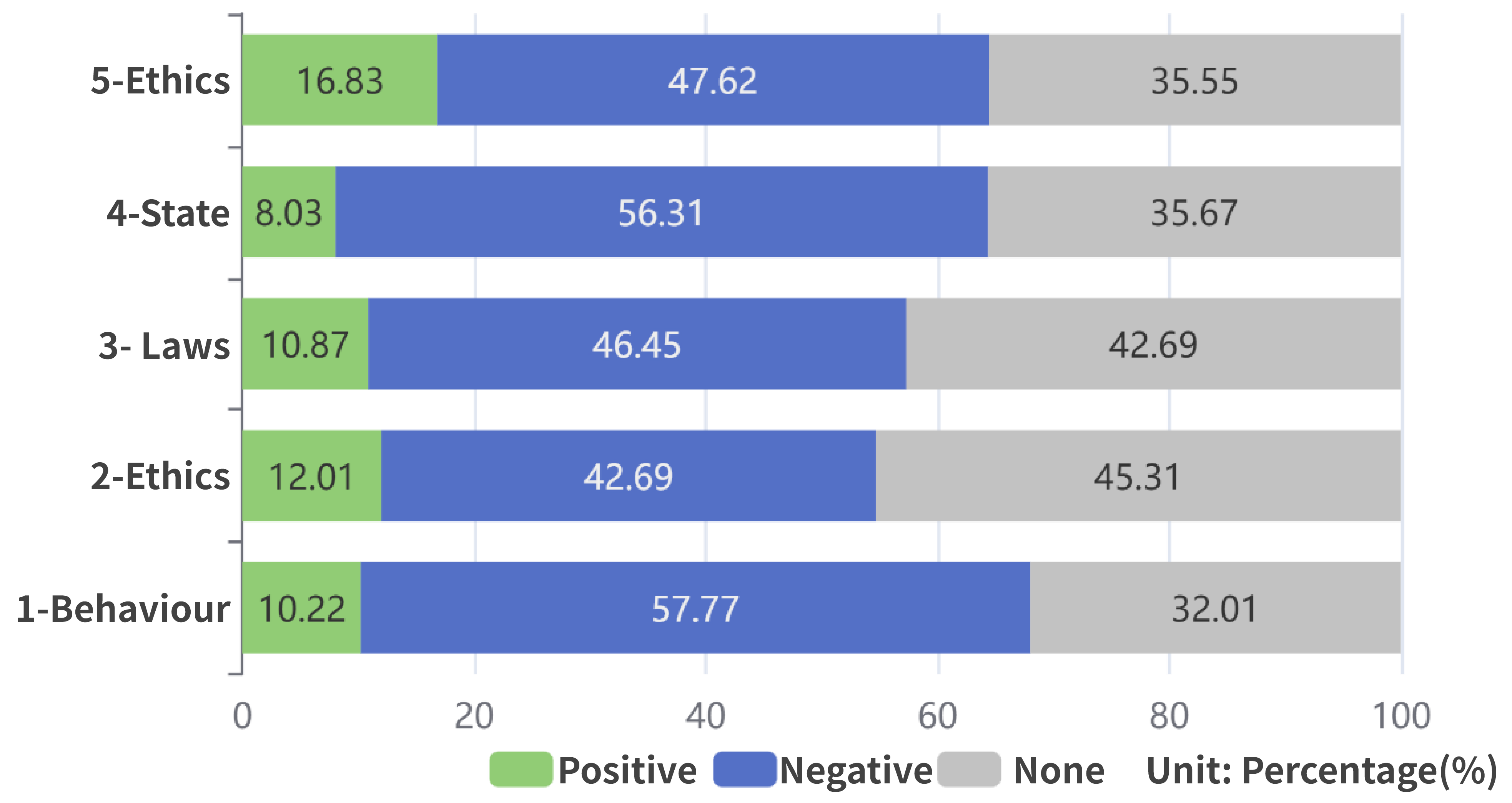}
		\caption{Results of Sentiment Share of Comments on Secondary Topics}
	\end{figure}
	
	\subsubsection{Individual dimension}
	Under Topic 1, some members of the public are hesitant to start a family, and are resistant to marriage and childbearing in the face of the economic downturn. Some opinion leaders have misinterpreted the respect for individual rights and interests and the basic protection of reproductive rights in their blog posts as blasphemy against the Marriage Law, and distorted them into proposals such as \textit{``encouraging illegitimate children"} and \textit{``supporting polygamy"}, which are very different from the traditional concepts of family morality, thus creating a significant emotional mobilisation effect and causing negative public opinion. This has had a significant emotional mobilising effect, generating negative public opinion and a markedly negative attitude among the public.
	
	In Topic 5, some members of the public are more concerned about issues such as \textit{``who will raise the child"}, \textit{``the child's future education and medical care"}, \textit{``whether the child has the right to inherit the property of the biological parents"}, etc. They also asked the media and relevant departments about the basic rights of citizens. The media and relevant departments have also been asked questions about the basic rights of citizens. At the same time, some groups find it difficult to accept social phenomena such as \textit{``single parenthood"}, and thus tend to misunderstand the starting point of the proposal. For example, the policy of simplifying the birth registration process, which respects women's freedom to give birth, has been mistaken for \textit{``preferential treatment for illegitimate children"}, or even that \textit{``The fermentation of these views has challenged the bottom line of family ethics and morality, and has led to negative public sentiments."}
	
	\subsubsection{Social dimension}
	The public's negative feelings on Topic2-Social Morality and Ethics are mainly due to the conflict between two pairs of relationships, namely, ``the conflict between moral concepts and the bombardment of information" and ``the split between order and chaos within the female community".
	
	On the one hand, the social nature of human beings determines the morals and norms that have long been commonly followed in dealing with each other. Under the cultural pattern of Chinese moral concepts and high contexts, such concepts as \textit{``monogamy"}, \textit{``the third party should be condemned"}, \textit{``the legitimacy of legitimate children"} etc., naturally become and are affirmed by society. The \textit{``legitimacy of children born in wedlock"} and so on have naturally become concepts that have been affirmed by society. However, the proposal to abolish the marriage restriction in the birth registry is like a stone that breaks the previous social concepts, and the proposal, which has been rendered by many Internet extremists as \textit{``encouraging out-of-wedlock births}", is even more shocking in a short phrase. The rhetoric surrounding the meaning of marriage, the significance of polygamy, and the bottom line of social morality all reflect the public's bewilderment and questioning of the past in the face of a sudden and strong ``new idea" and long-held moral values.
	
	On the other hand, public discussion has focused mostly on women, with two distinctly different attitudes towards the group of children born out of wedlock and the group of children born in wedlock. In the discussion on the group of children born out of wedlock, there is support for the \textit{``freedom of procreation"}, \textit{``untying marriage"}, and \textit{``improving the rights and interests of single mothers"}, while the opposite is true for the group of children born out of wedlock, \textit{``the existing marriage system"}, and \textit{``the rights and interests of women born in wedlock"}. On the other hand, there are concerns about the existing marriage system and the rights and interests of children born in wedlock in the group of children born in wedlock, and hostile attitudes towards each other are likely to emerge in the discussion of these two groups, thus intensifying the conflicts between the groups and making the public's overall mood negative.
	
	The public's discussion of Topic3-Laws and Regulations includes two dimensions: the proposal itself and the discussion of the proposal with other laws and regulations, while \textit{``abolishing the marriage restriction in the birth registration"} directly involves the formulation of laws and regulations, so the public's concern and emotion will be magnified. Negative emotions come not only from the overlapping of new and old problems, but also from the deviation of the real situation from the ideal environment.
	
	In a modern society where information pervades daily life, people are often caught in a flood of new problems and old problems that still need to be solved. The \textit{``abolition of marriage restrictions in birth registration"} is like a new issue in the flood, catching people by surprise and at the same time reminding them naturally of previous issues that need to be solved, such as \textit{``legalisation of surrogacy"}, \textit{``legalisation of IVF(in vitro fertilization)"}, etc. When these issues are combined with the new ones, they will be brought to mind as well. When these problems are linked to new ones, more problems arise, and in the overlapping of these problems, the public becomes sceptical and negative towards the law-making itself.
	
	At the same time, all kinds of information is flooded with the public can easily see the loopholes in the real world of laws and regulations, when we see the ``Fengxian birth of eight children", ``hot star surrogacy" and other social events frequently appear in front of us, people will be on the existing laws and regulations to make and the proposal of the staff have a distrustful attitude, and like \textit{``birth registration cancellation of marriage restrictions"} this proposal is the need for calm and objective thinking and relative stability of the social environment. When people lose trust in society, a proposal that is supposed to protect their rights and interests will be interpreted as \textit{``promoting childbirth"} or \textit{``squeezing labour"}, and when the gap between the current social environment and the ideal environment of the proposal is too wide, it is more likely to cause public dissatisfaction.
	
	\subsubsection{National dimension}
	The main sources of negative sentiments on Topic4-National Policies are ``the proposal is considered to be a departure from existing policies" and ``the focus of the proposal is contrary to society". During the discussion of this topic, some members of the public interpreted the proposal to protect single mothers as a proposal to promote population growth that is contrary to the traditional morals and ethics of the society, and contrary to the existing marriage policy system; the public tended to think that the proposal is to provide a new legal path to bypass the existing marriage system, and the public further expressed the view that the proposal is put forward without in-depth investigations and thinking, and that it is \textit{``not listening to the voice of the people"}. Another section of the public targeted at the member who put forward the proposal, describing it as a \textit{``blind proposal"} and an \textit{``ill-timed proposal"}, which further escalated into personal attacks on the member. Against the backdrop of a global economic downturn after three years of epidemics, the public expressed the hope that the members would put forward development proposals on current livelihood issues, such as education and the economy, rather than on such issues as marriage and childbirth.
	
	\section{Summary and discussion}
	Through this research, it is found that the audience's discussion under this topic shows significant negative emotions, and the overall discussion topics are divided into three directions, and the secondary topics are subdivided into five aspects. The multi-granularity opinion analysis framework combining semantic co-occurrence network and Bertopic topic model proposed in this paper yields results that are sufficiently detailed and interpretable to provide a reference approach for opinion research.
	
	\subsection{ From the topic of discussion}
	Users of the topic of the discussion is more focused on the social law, the face of the topic of the public questions, the government needs to simultaneously promote the fairness of the rights and interests of the matrix to improve the laws and regulations, and at the same time should strengthen the interpretation of the topic to guide.
	
	\textbf{Promoting equity in the protection of rights and benefits.} The imbalance in the protection of women's rights and interests is an important factor impeding the implementation of the policy, which should ensure that women who give birth to children out of wedlock are treated equally with those who give birth to children in wedlock in terms of maternity insurance, maternity benefits and maternity assistance. The Civil Code provides that children born out of wedlock have the same rights and obligations as those born in wedlock, and local governments need to make humane improvements to their policies in accordance with the law, so as to promote the construction of a birth-friendly society.
	
	\textbf{Improving laws and regulations in a matrix manner.} Fertility-related issues are easily associated with black industries such as ``surrogacy", ``sperm trading" and ``human trafficking", so it is necessary to improve relevant laws and regulations in a targeted, multi-faceted and matrix manner. It is necessary to improve the relevant laws and regulations in a targeted, multifaceted and matrixed manner, so as to solve existing problems and prevent ethical risks, avoid unnecessary overlapping discussions between new and old issues, avoid the risk of public opinion, and maintain a stable social order.
	
	\textbf{Strengthening the dissemination and interpretation of policies.} The Government and members should pay attention to the interpretation of policies and proposals, so as to avoid the intention of the policy issues being deviated and misinterpreted in social media. At the same time, it is important to clarify the focus of the publicity on the policy issue, as the abolition of the marriage restriction is intended to protect the rights and interests of the people, not to encourage unmarried pregnancies. In addition, the details of the proposals on birth registration and marriage restrictions should be enriched to promote effective dissemination and discussion of the proposals.
	
	\subsection{From the mood of the discussion}
	Users' negative expressions were significant under all topics, with some users showing transgressive behaviours under individual topics and a tendency towards group polarisation under some microblog tweets, for example, users' negative opinions around cyber-violence against individual committee members under the topic of national policies, and the perceived encouragement of illegitimate births under the topic of personal behaviours. In the face of this phenomenon, the government needs to enhance its anticipation of public opinion events in the early stage, cultivate opinion leaders with mainstream values in the medium term, and strengthen the online environment in the long term.
	
	\textbf{Enhance the anticipation of public opinion events.} There is a certain pattern of development in the formation of public opinion on the Internet, and it is only by predicting the direction of public opinion in advance and making preparations to deal with it that public opinion can be guided in a timely manner when it is formed. Especially when public policy issues such as childbirth and marriage are released, they will inevitably cause public opinion to develop, and it is necessary to prepare for the possible direction of public opinion in advance, and turn passive responses into active attacks. Only by taking the ``initiative" in guiding public opinion can we better guide public opinion and ensure the healthy development of online public opinion.
	
	\textbf{Cultivating opinion leaders with mainstream values.} Internet public opinion is easily controlled by a small number of ``opinion leaders", and some ``opinion leaders" on the Internet intentionally disseminate information on issues such as childbirth and marriage that is different from mainstream values, such as childbirth anxiety and marriage anxiety. Therefore, while trying to get some ``opinion leaders" on the Internet to work for us, we need to cultivate ``opinion leaders" who are in line with the mainstream values, and guide the Internet public opinion through the ``opinion leaders" to strengthen the mainstream public opinion. We also need to cultivate opinion leaders who are in line with mainstream values and who can guide online public opinion through them. They can answer questions about proposals and other information, guide public opinion correctly, and create a positive, healthy and upward-looking public opinion environment.
	
	\textbf{Strengthening the Internet environment.} The Internet is a new field for the struggle for public opinion, in which there is no lack of voices that take advantage of information asymmetry and other means to mislead society and disturb the public opinion environment. Therefore, it is imperative to purify the Internet environment and strengthen the construction of the Internet public opinion environment. Pragmatic, comprehensive and fair reporting and a prudent attitude should be used to squeeze the living space of various noises and murmurs with positive propaganda and eliminate the negative influence of various wrong and reactionary views with positive voices. On topics related to childbirth and marriage, we have strengthened our efforts to channel and guide public opinion, and have done more to resolve conflicts and sort out emotions. In the case of gender antagonism and the spread of anxiety, it is necessary to detect it in a timely manner and guide it in a gradual and orderly manner.
	
	\bibliographystyle{ACM-Reference-Format}
	\bibliography{sample-sigplan}

\end{document}